\documentclass[aps,showpacs,preprint,footinbib,preprintnumbers]{revtex4-1}

\usepackage{graphicx}
\usepackage{epstopdf}
\usepackage{amsmath,amssymb}
\usepackage{booktabs}
\usepackage{mathrsfs}

\newcommand{\be}{\begin{eqnarray}}
\newcommand{\ee}{\end{eqnarray}}

\newcommand{\vslash}{{v\hspace{-5.4pt}/}}

\usepackage[normalem]{ulem}  
\usepackage[dvips]{color} 
\renewcommand\sout{\bgroup \color{red} \ULdepth=-.5ex \ULset}

\begin{document}

\title{Exotic baryons from a heavy meson and a nucleon \\ -- Positive parity states --}

\author{Yasuhiro Yamaguchi$^1$}
\author{Shunsuke Ohkoda$^1$}
\author{Shigehiro Yasui$^2$}
\author{Atsushi Hosaka$^1$}
\affiliation{$^1$Research Center for Nuclear Physics (RCNP), 
Osaka University, Ibaraki, Osaka, 567-0047, Japan}
\affiliation{$^2$KEK Theory Center, Institute of Particle and Nuclear
Studies, High Energy Accelerator Research Organization, 1-1, Oho,
Ibaraki, 305-0801, Japan}


\begin{abstract}
We study heavy baryons with exotic flavor quantum numbers 
formed by a  heavy  meson and a nucleon ($\bar DN$ and $BN$)
with positive parity.
 One pion
 exchange interaction, providing a tensor force, dominates as a long
 range force to bind the $\bar DN$ and $BN$ systems.
In the heavy quark mass limit, pseudoscalar meson and vector meson
 are degenerate and
the binding mechanism by the tensor force analogous to that in the
 nuclear systems becomes important.
  As a result, we obtain the $\bar DN$ and $BN$ resonant states in
 the $J^P=1/2^+$, $3/2^+$ and $5/2^+$ channels with $I=0$.
\end{abstract}
\maketitle

\section{Introduction}
The recent finding of the twin $Z_b$ resonances
near the $B\bar B^*$ and $B^* \bar B^* $ thresholds~\cite{Collaboration:2011gja,Voloshin:2011qa,Bondar:2011ev} 
has added a new evidence of the exotic states in addition to candidates
such as $f_{0}$, $a_{0}$ and $\Lambda(1405)$ in strangeness sector~\cite{Oller:1997ti,Oset:1997it,Hyodo:2007jq}, $X$,
$Y$ and $Z$ in charm and bottom sectors~\cite{Brambilla:2010cs,Voloshin:2007dx,Lee:2009hy}, implying hadronic composites or
molecules.  
The appearance of the states near the threshold is a necessary condition that 
the states can be interpreted as hadronic composites with
keeping their identities as constituent hadrons.
The mechanism of forming hadronic composites near the threshold
depends on the nature of the interaction among the constituent hadrons.  

In this respect, the one pion exchange potential 
is of great interest as one of the most important meson-exchange potentials between the two constituent hadrons~\cite{Tornqvist:2004qy,Cohen:2005bx,Yasui:2009bz,Yamaguchi:2011xb}.  
The one pion exchange naturally works
when
the constituent hadrons have non-zero isospin values.
We note that the existence of the pion is a robust consequence of spontaneous breaking of 
chiral symmetry~\cite{Nambu:1961tp}.
A unique feature of the one pion exchange potential is the
tensor force due to the pseudoscalar nature of the pion.
The tensor force mixes the states with different angular momentum, i.e. $L$ and $L\pm 2$.  
This causes a mixing of the different configurations in a hadronic state and thus
yields an attraction between the two constituent hadrons with lower $L$ state.
In fact, it is known that the tensor force is the leading mechanism of the binding of atomic 
nuclei~\cite{Ericson_Weise}.

The pion exchange is possible also for the heavy hadron systems 
containing heavy pseudoscalar meson $P=\bar{D}$, $B$ and heavy 
vector meson $P^\ast=\bar{D}^\ast$, $B^\ast$.  
The Yukawa vertices of $PP^\ast \pi$ and $P^\ast P^\ast \pi$ generate the pion exchange 
potential which becomes important especially when $P$ and $P^\ast$ mesons 
are degenerate in the heavy quark limit.  
Here we note that only $P$ meson cannot generate the pion exchange 
potential because the $PP\pi$ vertex is not allowed due to the parity invariance.
In the literatures, this idea has been tested and shown to 
be indeed the case for heavy quark systems~\cite{Tornqvist:2004qy,Cohen:2005bx,Yasui:2009bz,Yamaguchi:2011xb}. 
This does not seem to work, however, for light flavor sector in which the heavy quark symmetry is not a good symmetry~\cite{Yamaguchi:2011xb}.  

In this paper, we study manifestly exotic baryons formed by a
$\bar{D}$ or $B$ meson and a nucleon $N$, $\bar{D}N$ and $BN$, whose minimal quark content is $\bar Q qqqq$, 
where $Q$ and $q$ stand for heavy and light quarks, respectively~\cite{comment}.
In Refs.~\cite{Yasui:2009bz,Yamaguchi:2011xb}, 
the investigation was made for the {\it negative}
parity states 
where the bound and resonant states were discussed.
Here we perform the analyses for the {\it positive}
 parity states to complete 
the investigation.  

This paper is organized as follows.  
In section 2, we briefly summarize the interaction between a heavy
meson $\bar{D}$ or $B$ and a nucleon $N$
by following the prescription in Refs.~\cite{Yasui:2009bz,Yamaguchi:2011xb}. 
In section 3, we solve the Schr\"odinger equations numerically and
search the bound and resonant states in several quantum numbers $(I,J^{P})$.
Unlike the negative parity states, bound states are not found 
in the positive parity states but only resonances are.
The difference of the present results from the previous ones is discussed.  
In the final section, we summarize the present work and discuss 
some future directions.  

\section{Interactions}

\begin{table}[tbp]
\centering
\caption{\label{table_qnumbers} \small Various coupled channels for a 
given quantum number $J^P$ for the positive parity $P = +1$.  }
\vspace*{0.5cm}
{\small 
\begin{tabular}{ c  | c c c c}
\hline
$J^P$ &  \multicolumn{4}{c }{channels} \\
\hline
$1/2^+$ &$PN(^2P_{1/2})$&$P^\ast N(^2P_{1/2})$&$P^\ast N(^4P_{1/2})$& \\
$3/2^+$ & $PN(^2P_{3/2})$&$P^\ast N(^2P_{3/2})$&$P^\ast
	      N(^4P_{3/2})$&$P^\ast N(^4F_{3/2})$ \\
$5/2^+$&$PN(^2F_{5/2})$&$P^\ast N(^4P_{5/2})$&$P^\ast
	      N(^2F_{5/2})$&$P^\ast N(^4F_{5/2})$ \\
\hline
\end{tabular}
}
\end{table}

Let us consider two-body states of a heavy meson and a nucleon with
positive parity. Those states can be classified by isospin $I$, total
spin $J$ and
parity $P$.
In the present study, we consider the states with $I=0$ and $1$, and $J^{P}=1/2^{+}$,
$3/2^{+}$ and $5/2^{+}$ , as summarized in
Table~\ref{table_qnumbers}.
In these systems, a heavy pseudoscalar meson ($P$) and a heavy
vector meson ($P^\ast$) are degenerate in the heavy quark limit.
Therefore, each state may contain both $P$ and $P^\ast$, leading to
a problem with coupled channels; three channels for $J^{P}=1/2^{+}$ and four channels
for $J^{P}=3/2^{+}$ and $5/2^{+}$.

To obtain the interactions for a heavy meson and a nucleon, we employ Lagrangians 
satisfying the heavy quark symmetry and chiral symmetry \cite{Manohar:2000dt,Casalbuoni}.  
They are well-known and given as
\be
{\cal L}_{\pi HH} &=&   ig_\pi \mbox{Tr} \left[
H_b\gamma_\mu\gamma_5 A^\mu_{ba}\bar{H}_a \right]   \, , 
\label{LpiHH}\\
{\cal L}_{v HH} &=&  -i\beta\mbox{Tr} \left[ H_b
v^\mu(\rho_\mu)_{ba}\bar{H}_a \right]
+i\lambda\mbox{Tr} \left[
H_b\sigma^{\mu\nu}F_{\mu\nu}(\rho)_{ba}\bar{H}_a \right] \, , 
\label{LvHH}
\ee
where the subscripts $\pi$ and $v$ ($=\rho$ and $\omega$) are for the pion 
and vector mesons.
We consider not only the pion exchange which is
relevant at long distances, but also the vector meson exchange which is
relevant at short distances.
In Eq.~\eqref{LvHH}, $v^{\mu}$ is the four-velocity of a heavy quark.
In Eqs.~\eqref{LpiHH} and \eqref{LvHH}, the heavy meson fields of $\bar Q q$ are parametrized by 
the heavy pseudoscalar and vector mesons, 
\be
H_a   &=& \frac{1+\vslash}{2}\left[P^\ast_{a\,\mu}\gamma^\mu-P_a\gamma_5\right] \, ,  \\
\bar H_a  &=& \gamma_0 H^\dagger_a \gamma_0 \, ,
\ee
where the subscripts $a$, $b$ are for light flavors, $u$, $d$.  
From Eqs.~\eqref{LpiHH} and \eqref{LvHH}, we obtain the pion and vector
meson vertices in the static approximation $v^{\mu} = (1, \vec{0})$.
The coupling constants $g_\pi$, $\beta$ and $\lambda$ are the same as in
our previous papers Refs.~\cite{Yasui:2009bz,Yamaguchi:2011xb} as
summarized in Table~\ref{table_Nconstants}.

The interaction Lagrangians for a meson and nucleons are given by the standard form, 
\be
{\cal L}_{\pi NN} &=& \sqrt{2} ig_{\pi
NN}\bar{N}\gamma_5 \hat{\pi} N \, , \label{LpiNN} \\
{\cal L}_{vNN} &=& \sqrt{2} g_{vNN}\left[\bar{N} \gamma_\mu \hat{\rho}^\mu N 
+\frac{\kappa}{2m_N}\bar{N} \sigma_{\mu\nu} \partial^\nu \hat{\rho}^\mu
N \right] \, , \label{LvNN}
\ee
where $N =(p,n)^T$ is the nucleon field.
The coupling constants for the nucleon are taken from the phenomenological nuclear potential in Ref.~\cite{machleidt} 
as summarized in Table~\ref{table_Nconstants}.

\begin{table}[tbp]
\centering
\caption{\label{table_Nconstants} Masses and coupling constants of
 mesons in Ref.~\cite{Yamaguchi:2011xb}.}
\vspace*{0.5cm}
{\small 
\begin{tabular}{ c  c c c c c c}\hline
  & $m_{\alpha}\,\,[\mbox{MeV}]$ & 
 $g_\pi$ & $\beta$ & $\lambda\,\,[\mbox{GeV}^{-1}]$&  $g^2_{\alpha NN}/4\pi$ & $\kappa$  \\
\hline
$\pi$ & 137.27 & 0.59 & --- & ---  &  13.6 & ---  \\
$\rho$ & 769.9 & --- & 0.9 & 0.56 & 0.84 & 6.1 \\
$\omega$ & 781.94 & --- & 0.9 & 0.56 & 20.0 & 0.0 \\ \hline
\end{tabular}
}
\end{table}

\begin{table}
  \begin{center}
\caption{Cut-off parameters of a nucleon and heavy mesons in
   Ref.~\cite{Yamaguchi:2011xb}.}
\label{cutoff}
   \begin{tabular}{cccc}
\hline
   Potential &$\Lambda_N$ [MeV] &$\Lambda_D$ [MeV] &$\Lambda_B$ [MeV] \\
\hline
   $\pi$&830&1121&1070 \\
   $\pi \rho\, \omega$&846&1142&1091 \\
\hline
   \end{tabular}
  \end{center}
 \end{table}

The potentials are derived by the vertices~\eqref{LpiHH}, \eqref{LvHH}, \eqref{LpiNN} and \eqref{LvNN}
as shown in Appendix.
To take into account the internal structure of the hadrons, form factors associated 
with finite size of the mesons and nucleons are introduced at each vertex.
We introduce the monopole type form factors as shown in Appendix.
Here we have two cut-off parameters for $\bar{D}$ ($B$) mesons and a nucleon.
The cut-off parameter for the nucleon is determined to reproduce the
properties of the deuteron: the binding energy, scattering length and
effective range.
The cut-off parameters for $\bar{D}$ ($B$) mesons are determined from the
ratios of matter radii of $\bar{D}$ ($B$) meson and nucleon, which are estimated
by a quark model, as discussed in
Ref.~\cite{Yasui:2009bz,Yamaguchi:2011xb}.
Using those potentials, we solve the coupled-channel Schr\"odinger
equations numerically.
We employ the two potentials; the $\pi$ exchange potential and the $\pi
\rho\, \omega$ potential to discuss the important role of the pion.
The cut-off parameters of the nucleon vertices for each parameter set
are 
summarized in
Table~\ref{cutoff}.

\section{Scattering states and resonances}

After solving the Schr\"odinger equations,
we find no bound state
neither for $\bar{D}N$ nor for $BN$ systems.
However, by analyzing the scattering states, we find several resonances
in the isosinglet channel ($I=0$) both for $\bar{D}N$ and for $BN$, but no structure in the
isotriplet channel ($I=1$).

Let us first explain how resonant states are determined in the present study. The resonance energy and decay width
are obtained by analyzing the phase shift in the scattering
states.
In the previous work~\cite{Yamaguchi:2011xb}, we
identified a
resonance at the position where the phase shift crosses $\pi/2$, because
the decay width was relatively smaller than the resonance energy
from the threshold.
In the present study, because it will turn out that the decay widths are not
necessarily small,
we define the resonance position $E_{\mathrm{re}}$ by an inflection point of the
phase shift~\cite{Arai:1999pg}. Then, the width is obtained by
$\Gamma=2/(d\delta/dE)_{E=E_{\mathrm{re}}}$ at the inflection point.

\begin{figure}[htbp]
 \begin{center}
  \begin{tabular}{cc}
{\includegraphics[width=80mm,clip]{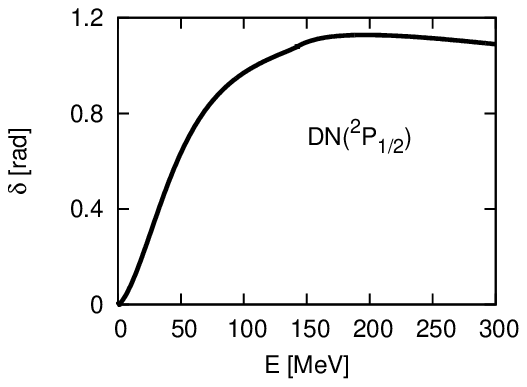}} &
{\includegraphics[width=80mm,clip]{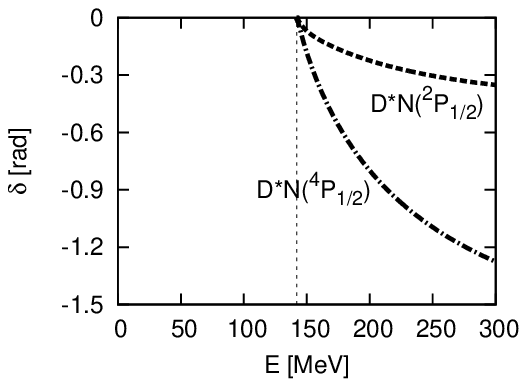}}
\\
{\includegraphics[width=80mm,clip]{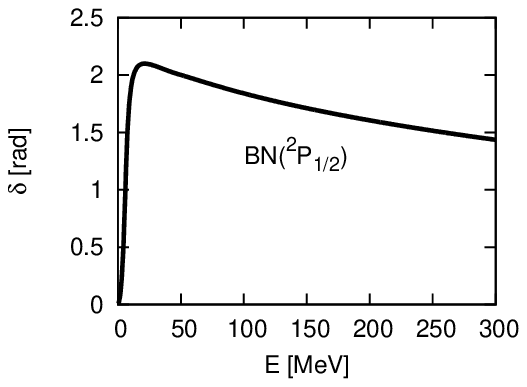}} &
{\includegraphics[width=80mm,clip]{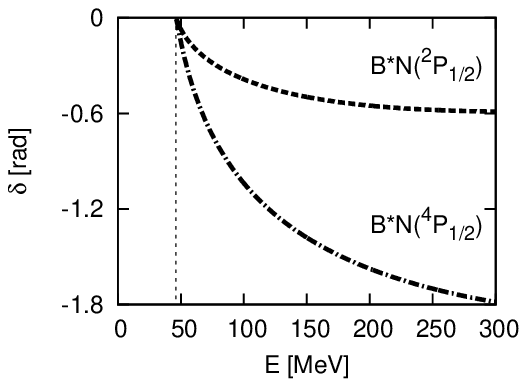}}
\\
\end{tabular}  
\caption{Phase shifts of the $\bar{D}N$ and $BN$ scattering states with
  $(I,J^P)=(0,1/2^+)$  when the $\pi\rho\,\omega$ potential is used.}
\label{phaseBN1/2}
 \end{center}
\end{figure}

\begin{figure}[htbp]
 \begin{center}
  \begin{tabular}{cc}
   \begin{minipage}{.5\textwidth}
    {\includegraphics[width=80mm,clip]{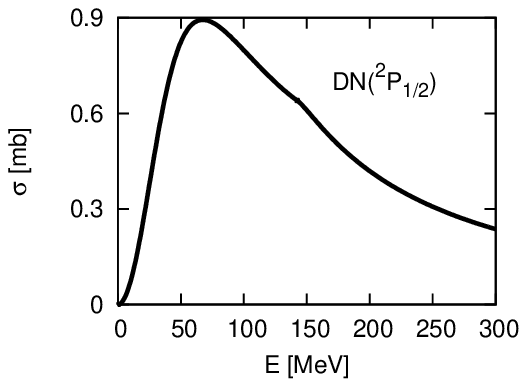}}
\caption{Total cross section of the $\bar{D}N$ scattering state with
  $(I,J^P)=(0,1/2^+)$  when the $\pi\rho\,\omega$ potential is used.}
\label{crossDN1/2}
   \end{minipage}
   &\quad
   \begin{minipage}{.5\textwidth}
    {\includegraphics[width=80mm,clip]{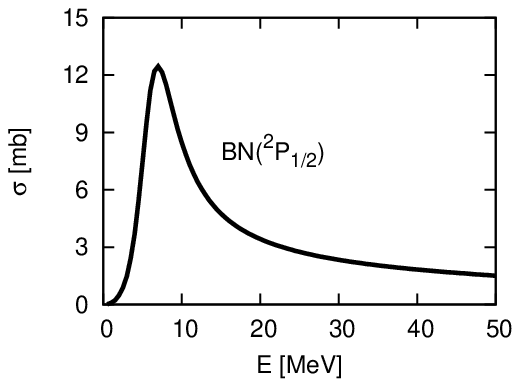}}
\caption{Total cross section of the $BN$ scattering state with
  $(I,J^P)=(0,1/2^+)$  when the $\pi\rho\,\omega$ potential is used.}
\label{crossBN1/2}
   \end{minipage}
 \\
  \end{tabular}
 \end{center}
\end{figure}

Now we discuss the result of each channel $(I,J^{P})=(0,1/2^{+})$, $(0,3/2^{+})$ and $(0,5/2^{+})$.
In the $(I,J^P)=(0,1/2^+)$ channel, we find resonances in both
$\bar{D}N$ and
$BN$ systems.
The phase shifts $\delta$'s of $PN(^2P_{1/2})$, $P^\ast
N(^2P_{1/2})$ and $P^\ast N(^4P_{1/2})$ channels obtained for the
$\pi \rho \, \omega$ potential are shown as functions of the
scattering energy $E$ in the center of mass system in
Fig.~\ref{phaseBN1/2}.
The vertical dashed lines in the figures represent the positions of $\bar{D}^\ast N$ and
$B^\ast N$ thresholds.
The resonance energies are measured from the lowest thresholds
($\bar{D}N$ and $BN$).
The sharp increase of the phase shift of the $PN(^2P_{1/2})$ channel
indicates the existence of a resonance.
A similar behavior is obtained also when the $\pi$ exchange potential is employed.
The resonance energies
and decay widths are summarized in Table~\ref{table4}.
We obtain the resonance energy at 26.8 MeV for
$\bar{D}N$ and at 5.8 MeV
for $BN$, with the decay widths 131.3 MeV and 
6.0 MeV, respectively, for the $\pi \rho\, \omega$ potential.
Here, we compare two results by the $\pi$ exchange potential and 
the $\pi\rho\,\omega$ potential, and find the difference is very small.
 As a result,
the vector meson ($\rho$ and $\omega$) exchange interaction plays a minor role, while
the $\pi$ exchange interaction plays a dominant role to generate
resonant states.
In particular, the tensor force is important because no bound or resonant state exists without
the $PN$-$P^\ast N$ mixing.
The pion dominance was also seen for the negative party states as in our previous work~\cite{Yamaguchi:2011xb}.
The resonances are generated below the $P^{\ast}N$ threshold, where the $PN$ channel is open and the $P^{\ast}N$ channel is closed.
Here, we note that the $PN$-$PN$ channel has no interaction in the pion
exchange potential due to the parity non-conservation at the $PP\pi$
vertex, as presented explicitly in the $\pi$ exchange potential (\ref{matpi}) in Appendix.
Therefore the attractive force which forms the resonance in the $PN$ channel is mainly provided from the $PN$-$P^{\ast}N$ mixing effect.
As a consequence, the mixing effect yields sufficient attraction to form
the so-called shape resonance in the $PN(^{2}P_{1/2})$ channel with the p-wave
centrifugal barrier.
The total cross sections for the $\bar{D}N$ and $BN$ scatterings when the $\pi \rho \, \omega$
potential is used are shown in
Figs.~\ref{crossDN1/2} and~\ref{crossBN1/2}, respectively.
The peaks are found at around each resonance energy, 26.8 MeV and 5.8 MeV, for $\bar{D}N$ and $BN$, respectively.

\begin{table}[h]
\begin{tabular}{c}
\begin{minipage}{\textwidth}
\caption{The resonance energies $E_{\mathrm{re}}$ and decay widths $\Gamma$ for $(I,J^P)=(0,1/2^+)$.}
\label{table4}
 \begin{center}
  \begin{tabular}{lcccc}\hline 
&$\bar{D}N(\pi)$&$\bar{D}N(\pi\rho\,\omega)$&$BN(\pi)$&$BN(\pi\rho\,\omega)$ \\
\hline
$E_{\mathrm{re}}$ [MeV]&26.1&26.8&5.8&5.8 \\
$\Gamma$ [MeV]&125.2&131.3&5.8&6.0 \\
\hline
  \end{tabular}
 \end{center}
 \end{minipage}
\end{tabular}
\end{table}


\begin{figure}[htbp]
 \begin{center}
  \begin{tabular}{cc}
 {\includegraphics[width=80mm,clip]{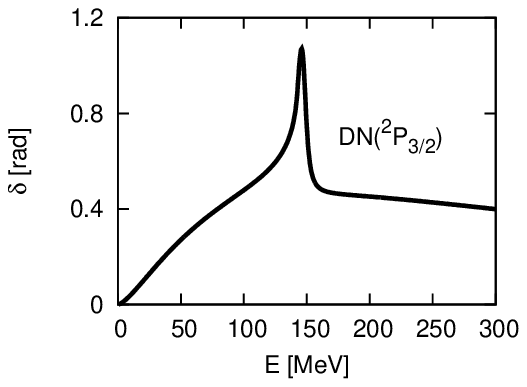}} &
 {\includegraphics[width=80mm,clip]{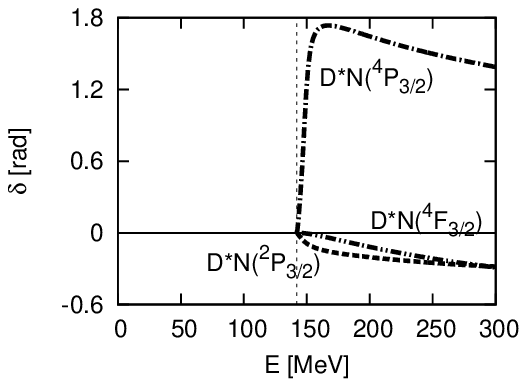}}
 \\
 \end{tabular}  
\caption{Phase shifts of the $\bar{D}N$ scattering states with
  $(I,J^P)=(0,3/2^+)$  when the $\pi\rho\,\omega$ potential is used.}
\label{phaseDN3/2}
 \end{center}
\end{figure}

\begin{figure}[htbp]
 \begin{center}
\begin{tabular}{cc}
\begin{minipage}{.5\textwidth}
 {\includegraphics[width=80mm,clip]{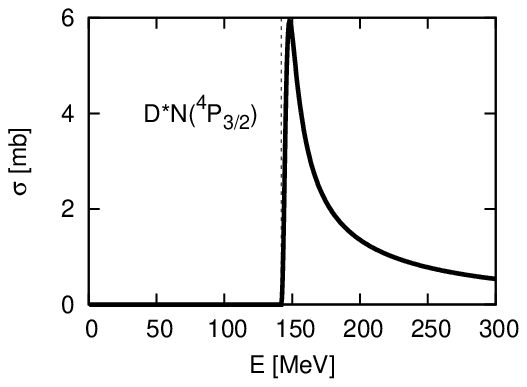}}
\caption{Total cross section of the $\bar{D}^\ast N(^4P_{3/2})$ scattering
when the $\pi\rho\,\omega$ potential is used.}
\label{crossDN3/2}
\end{minipage}
 &\quad
\begin{minipage}{.5\textwidth}
{\includegraphics[width=80mm,clip]{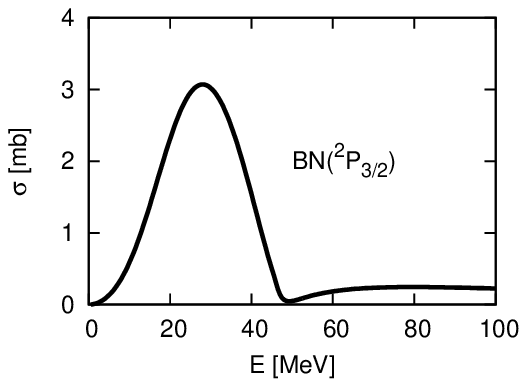}}
\caption{Total cross section of the $BN$ scattering state with
  $(I,J^P)=(0,3/2^+)$  when the $\pi\rho\,\omega$ potential is used.}
\label{crossBN3/2}
\end{minipage}
 \\
\end{tabular}
 \end{center}
\end{figure}

In the $(I,J^P)=(0,3/2^+)$ channel, we find a resonance for each
$\bar{D}N$ and $BN$ state. For $\bar{D}N$ state,
we show the phase shifts $\delta$'s of
$\bar{D}N(^2P_{3/2})$, $\bar{D}^{\ast}N(^2P_{3/2})$,
$\bar{D}^{\ast}N(^4P_{3/2})$ and $\bar{D}^{\ast}N(^4F_{3/2})$ in
Fig.~\ref{phaseDN3/2}. 
There is a small peak structure in the phase shift of
$\bar{D}N(^2P_{3/2})$ at the $\bar{D}^\ast N$ threshold, which is
interpreted as a cusp.
On the other hand, the phase shift of $\bar{D}^\ast
N(^4P_{3/2})$ which rises sharply indicates the presence of a
resonance in this channel. 
Therefore the resonant state exists
in the $\bar{D}^\ast N(^4P_{3/2})$ channel.
The resonance energies and decay widths are summarized in Table~\ref{table5}.
When the $PN$-$P^{\ast}N$ mixing is ignored, there still exists a resonance at the resonance
energy 145.5 MeV with the decay width 6.1 MeV, which are close to the original values in the full channel-couplings.
Therefore, the obtained resonance is a shape resonance generated mainly
by the p-wave centrifugal barrier in the $\bar{D}^{\ast}N(^{4}P_{3/2})$
states in the $P^{\ast}N$ channel.
We show the cross section for $\bar{D}N(^4P_{3/2})$ for the
$\pi\rho\,\omega$ potential in Fig.~\ref{crossDN3/2}, in which the peak appears at the resonance energy 148.2 MeV.

\begin{figure}[htbp]
 \begin{center}
  \begin{tabular}{cc}
{\includegraphics[width=80mm,clip]{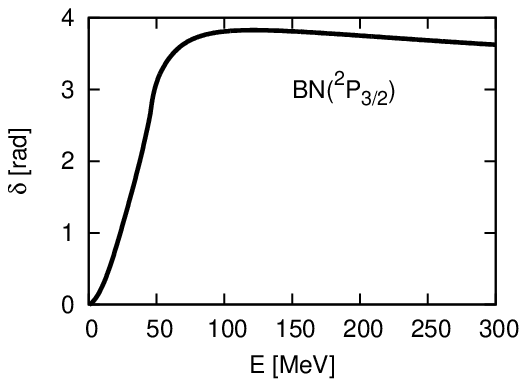}} &
{\includegraphics[width=80mm,clip]{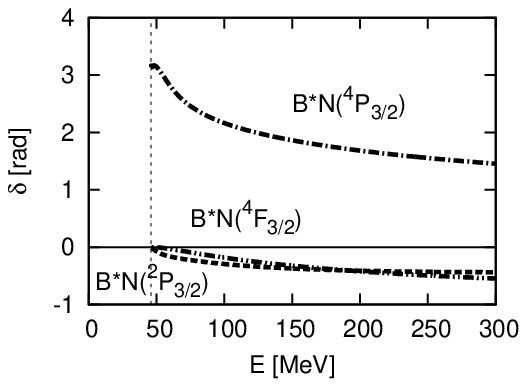}}
\\
\end{tabular}  
\caption{Phase shifts of the $BN$ scattering states with
  $(I,J^P)=(0,3/2^+)$  when the $\pi\rho\,\omega$ potential is used.}
\label{phaseBN3/2}
 \end{center}
\end{figure}

For the $BN$ state with $(I,J^P)=(0,3/2^+)$,
the phase sfhifts $\delta$'s of $BN(^2P_{3/2})$, $B^{\ast}N(^2P_{3/2})$,
$B^{\ast}N(^4P_{3/2})$ and $B^{\ast}N(^4F_{3/2})$ are plotted in
Fig.~\ref{phaseBN3/2}.
We find that the sharp increase of the phase shift passing through
$\pi/2$ in $BN(^2P_{3/2})$ as an indication of a resonance.
We also find that the phase shift in $B^{\ast}N(^4P_{3/2})$ starts from $\pi$.
Therefore, the obtained resonance can be regarded as a bound state of $B^{\ast}N$.
Indeed, when we switch off the $BN(^2P_{3/2})$ channel, we obtain a bound state of $B^{\ast}N$ which energy is close to the original resonance position.
Therefore, we conclude that the resonance in the $BN$ state with $(I,J^P)=(0,3/2^+)$ is a Feshbach resonance.
The resonance energy is 31.8 MeV and the decay width is 28.7 MeV as summarized in Table~\ref{table5}.
In Fig.~\ref{crossBN3/2}, we plot the cross section for $BN$ with
$(I,J^P)=(0,3/2^+)$, where we see a peak at around the resonance energy 31.8 MeV.

\begin{table}[h]
\caption{The resonance energies $E_{\mathrm{re}}$ and decay widths $\Gamma$ for $(I,J^P)=(0,3/2^+)$.}
\label{table5}
 \begin{center}
  \begin{tabular}{lcccc}\hline 
&$\bar{D}^\ast N(^4P_{3/2})(\pi)$&$\bar{D}^\ast N(^4P_{3/2})(\pi\rho\,\omega)$&$BN(\pi)$&$BN(\pi\rho\,\omega)$ \\
\hline
$E_{\mathrm{re}}$ [MeV]&148.2&148.2&32.3&31.8 \\
$\Gamma$ [MeV]&10.0&10.1&28.9&28.7 \\
\hline
  \end{tabular}
 \end{center}
\end{table}

\begin{figure}[htbp]
 \begin{center}
  \begin{tabular}{cc}
 {\includegraphics[width=80mm,clip]{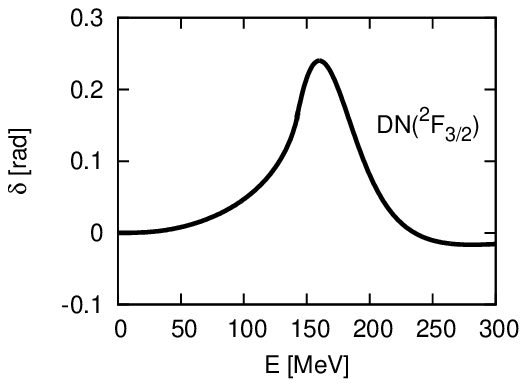}} &
 {\includegraphics[width=80mm,clip]{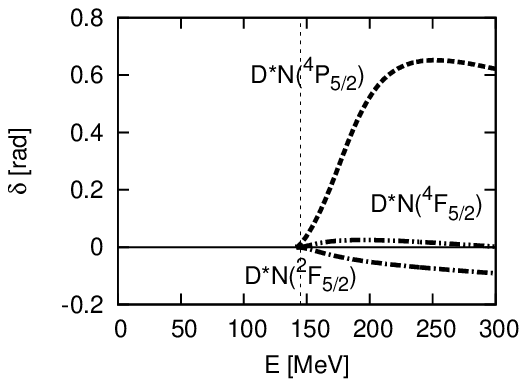}}
 \\
 {\includegraphics[width=80mm,clip]{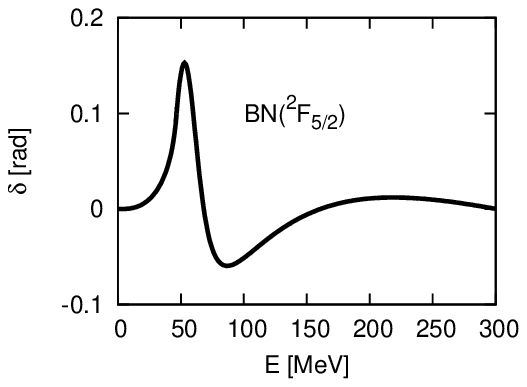}} &
 {\includegraphics[width=80mm,clip]{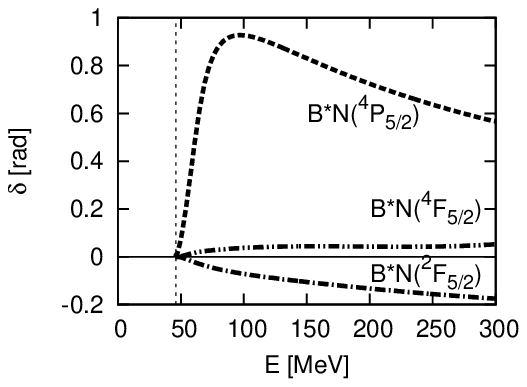}}
 \\
 \end{tabular}  
\caption{Phase shifts of the $\bar{D}N$ and $BN$ scattering states with
  $(I,J^P)=(0,5/2^+)$ when the $\pi\rho\,\omega$ potential is used.}
\label{phaseDN5/2}
 \end{center}
\end{figure}

\begin{figure}[htbp]
 \begin{center}
   \begin{tabular}{cc}
    \begin{minipage}{.5\textwidth}
   {\includegraphics[width=80mm,clip]{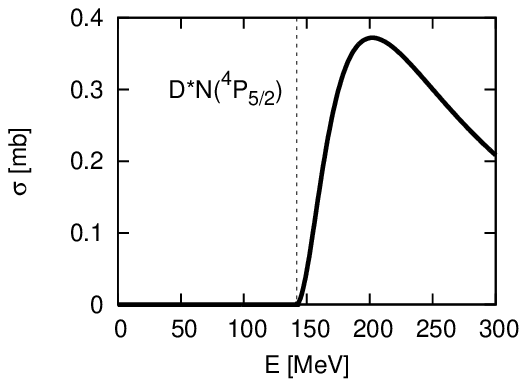}}
\caption{Total cross section of the $\bar{D}^\ast N(^4P_{5/2})$ scattering
when the $\pi\rho\,\omega$ potential is used.}
\label{crossDN5/2}
  \end{minipage}
&\quad
 \begin{minipage}{.5\textwidth}
   {\includegraphics[width=80mm,clip]{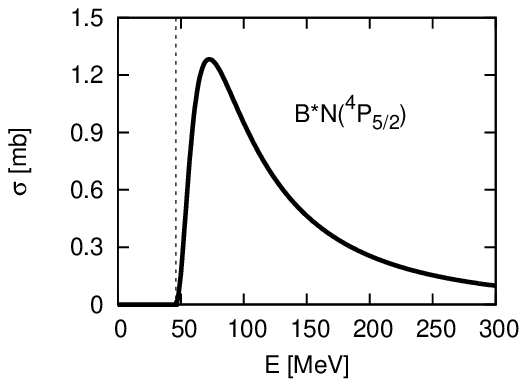}}
\caption{Total cross section of the $B^\ast N(^4P_{5/2})$ scattering
when the $\pi\rho\,\omega$ potential is used.}
\label{crossBN5/2}
  \end{minipage}
 \\
   \end{tabular}
 \end{center}
\end{figure}

In the $(I,J^{P})=(0,5/2^{+})$ channel, we find a resonance for each
$\bar{D}N$ and $BN$ above the $P^\ast N$ threshold. The phase shifts $\delta$'s
of $PN(^2F_{5/2})$, $P^\ast N(^4P_{5/2})$, $P^\ast N(^2F_{5/2})$ and
$P^\ast N(^4F_{5/2})$ are shown in Fig.~\ref{phaseDN5/2}. 
Small peak structures in the phase shifts of $\bar{D}N(^2F_{5/2})$ and
$BN(^2F_{5/2})$ are interpreted as cusps.
Above the $P^\ast N$ threshold, the phase shifts of $\bar{D}^\ast N(^4P_{5/2})$
and $B^\ast N(^4P_{5/2})$ rise up and these structures indicate the
presence of resonances. The resonance energies are 176.0 MeV for
$\bar{D}^\ast N$ and 58.4 MeV for $B^\ast N$, and the decay widths are
174.8 MeV and 49.6 MeV, respectively. We summarize the results in
Table~\ref{table5/2}. 
When the $PN$-$P^{\ast}N$ mixing is ignored,
the resonant states in the $\bar{D}^\ast N(^4P_{5/2})$
and $B^\ast N(^4P_{5/2})$ channels still exist at the resonance energies close to the values from the full channel-couplings.
Therefore, these resonant states are shape resonances generated mainly by the p-wave centrifugal barrier in the $P^{\ast}N(^{4}P_{5/2})$ channel.
The cross sections for $\bar{D}^\ast
N(^4P_{5/2})$ and $B^\ast N(^4P_{5/2})$ are plotted in
Figs.~\ref{crossDN5/2} and~\ref{crossBN5/2}, respectively.

Several comments are in order. First, in each $(I,J^{P})$ state, we verify
that the results of the $\pi$ exchange potential is similar to those of the $\pi
\rho \, \omega$ potential.
It means that both $\bar{D}N$ and $BN$ systems are dominated almost by the long
range force due to the pion exchange.
Second, the mixing between $PN$ and $P^{*}N$ is important for the
positive parity states, as for the negative parity states~\cite{Yasui:2009bz,Yamaguchi:2011xb}.
The $\bar{D}N$ and $BN$ resonances with $(I,J^{P})=(0,1/2^+)$ are
generated with the p-wave centrifugal barrier in the $PN$ channel by the attraction induced from the $PN$-$P^\ast N$ mixing.
The $PN$-$P^\ast N$ mixing is important also for the $BN$ resonant state with $(I,J^{P})=(0,3/2^{+})$ because it is a Feshbach resonance.
However, the $\bar{D}N$ resonance with $(I,J^{P})=(0,3/2^+)$ and the
$\bar{D}N$ and $BN$ resonances with $(I,J^{P})=(0,5/2^+)$ are generated
with the p-wave centrifugal barrier in the $P^{\ast}N$ channel mainly by the $P^{\ast}N$ interaction, in which the $PN$-$P^{\ast}N$ mixing effect plays a minor role.
Therefore, in the positive parity states, resonances are generated by
different mechanisms as summarized in Table \ref{resonance_mechanisms}.
We note that the Feshbach resonance in the negative parity states was obtained both for $\bar{D}N$ and $BN$ states with $(I,J^{P})=(0,3/2^{-})$ in Ref.~\cite{Yamaguchi:2011xb}.

\begin{figure}[htbp]
 \begin{center}
  {\includegraphics[width=170
mm,clip]{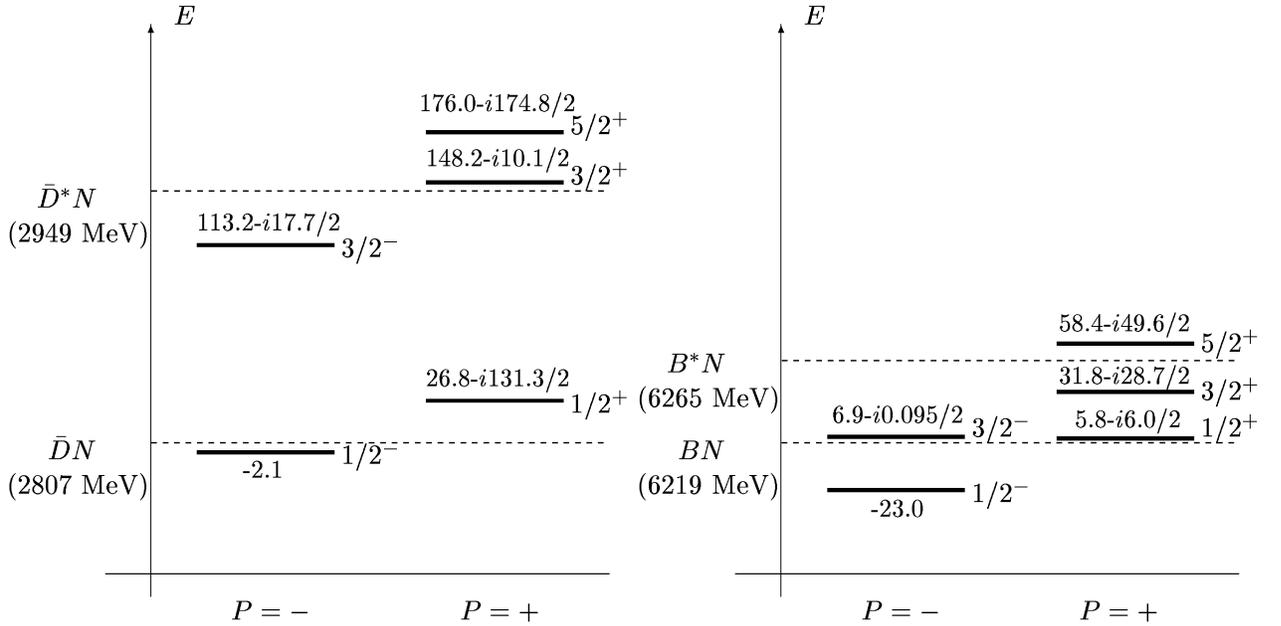}}
\caption{Exotic states with positive parity ($P=+$) and
 negative parity ($P=-$). The energies are measured from the lowest
  thresholds ($\bar{D}N$ and $BN$). The binding energy is given as a
  real negative value, and the resonance energy $E_{\mathrm{re}}$ and
  decay width $\Gamma$ are given as $E_{\mathrm{re}}-i\Gamma/2$, in
  units of MeV. The values are given when the $\pi\rho\,\omega$
  potential is used.}
\label{energylevel}
 \end{center}
\end{figure}

To compare the result of the positive parity states
with the result of the negative parity states \cite{Yasui:2009bz,Yamaguchi:2011xb},
we show energy levels for the exotic states found in our
investigations
in Fig.~\ref{energylevel}. 
We find that the bound states exist in the negative parity states,
while no bound state exists in the positive parity states.
This is because the lowest state $PN(^2P_{1/2})$ in $J^P=1/2^+$ has
a p-wave orbital angular momentum
$L=1$, while $PN(^2S_{1/2})$ in $J^P=1/2^-$ has an s-wave orbital angular momentum $L=0$.
In the same way, we see that the resonance energies of the positive parity states tend to be higher
than those of the negative parity states.

\begin{table}[h]
\caption{The resonance energies $E_{\mathrm{re}}$ and decay widths $\Gamma$ for $(I,J^P)=(0,5/2^+)$.}
\label{table5/2}
 \begin{center}
  \begin{tabular}{lcccc}\hline 
&$\bar{D}^\ast N(^4P_{5/2})(\pi)$&$\bar{D}^\ast
   N(^4P_{5/2})(\pi\rho\,\omega)$&$B^\ast N(^4P_{5/2})(\pi)$&$B^\ast N(^4P_{5/2})(\pi\rho\,\omega)$ \\
\hline
$E_{\mathrm{re}}$ [MeV]&177.1&176.0&58.5&58.4 \\
$\Gamma$ [MeV]&184.6&174.8&52.2&49.6 \\
\hline
  \end{tabular}
 \end{center}
\end{table}


\begin{table}[htdp]
\caption{The mechanism to form the resonances in the $\bar{D}N$ and $BN$
 states with $(I,J^{P})=(0,1/2^{+})$, $(0,3/2^{+})$ and
 $(0,5/2^{+})$. All the shape resonances are induced by the p-wave
 centrifugal barrier in the $PN$ and $P^{\ast}N$ channels,
 respectively.}
\begin{center}
\begin{tabular}{c|c|c}
\hline
$(I,J^{P})$ & $\bar{D}N$ states & $BN$ states \\
\hline \hline
$(0,1/2^{+})$ & \multicolumn{2}{c}{shape resonance in $PN$} \\
\hline
$(0,3/2^{+})$ & shape resonance in $P^{\ast}N$ & Feshbach resonance \\
\hline
$(0,5/2^{+})$ & \multicolumn{2}{c}{shape resonance in $P^{\ast}N$} \\
\hline
\end{tabular}
\end{center}
\label{resonance_mechanisms}
\end{table}%

\clearpage
\section{Summary}

We have investigated exotic baryons constructed by a heavy
meson and a nucleon like $\bar{D}N$ and $BN$.
The interaction is given by the meson exchange potential between a
$P=\bar{D}, B$ meson and a nucleon $N$, with respecting the heavy quark
symmetry.
To form those resonances, the $PN$-$P^\ast N$
mixing originating from the heavy quark symmetry plays a crucial role.
We have used the $\pi$ exchange potential and the $\pi \rho\,\omega$
potential, and found that the pion exchange interaction works dominantly,
while
the vector meson exchange interaction plays only a minor role.
Unlike the previous study for the negative parity case
\cite{Yasui:2009bz,Yamaguchi:2011xb}, there is no bound state in the
positive
parity case.
However,
we have found new resonances for $(I,J^P)=(0,1/2^+)$,
$(0,3/2^+)$ and $(0,5/2^+)$ in isosinglet states, and no structure in isotriplet states.
These resonant states exist near or below the thresholds of $P^{*}N$.
These exotic systems will be
interesting objects which can be searched and studied at accelerator facilities,
such as J-PARC, FAIR and so on, and also in the relativistic heavy ion collisions at RHIC and LHC \cite{Cho:2010db,Cho:2011ew}.

\clearpage
\section*{Acknowledgments}
This work is supported in part by a Grant-in-Aid for Scientific Research on 
Priority Areas ``Elucidation of New Hadrons with a Variety of Flavors 
(E01: 21105006)" from 
the ministry of Education, Culture, Sports, Science and Technology of Japan.

\appendix
\section{Potentials and kinetic terms}
\label{appendix_a}

The interaction potentials are derived by using the Lagrangians 
Eqs.~\eqref{LpiHH}-\eqref{LvNN}. In deriving the potentials we use
the static approximation where the energy transfer can be
neglected as compared to the momentum transfer.
The resulting potentials for the coupled channel systems are 
given in the matrix form of $3 \times 3$ for $J^{P}=1/2^+$ and
of $4 \times 4$ for $J^{P}=3/2^+$ and $5/2^+$,
 
\begin{eqnarray}
& V_{1/2^+} =
\begin{pmatrix}
V^{11}_{1/2^+} & V^{12}_{1/2^+} & V^{13}_{1/2^+} \\
V^{21}_{1/2^+} & V^{22}_{1/2^+} & V^{23}_{1/2^+} \\
V^{31}_{1/2^+} & V^{32}_{1/2^+} & V^{33}_{1/2^+}
\end{pmatrix}
\, , \\
& V_{3/2^+,5/2^+} =
\begin{pmatrix}
V^{11}_{3/2^+,5/2^+} & V^{12}_{3/2^+,5/2^+} & V^{13}_{3/2^+,5/2^+} & V^{14}_{3/2^+,5/2^+} \\
V^{21}_{3/2^+,5/2^+} & V^{22}_{3/2^+,5/2^+} & V^{23}_{3/2^+,5/2^+} & V^{24}_{3/2^+,5/2^+} \\
V^{31}_{3/2^+,5/2^+} & V^{32}_{3/2^+,5/2^+} & V^{33}_{3/2^+,5/2^+} & V^{34}_{3/2^+,5/2^+} \\
V^{41}_{3/2^+,5/2^+} & V^{42}_{3/2^+,5/2^+} & V^{43}_{3/2^+,5/2^+} & V^{44}_{3/2^+,5/2^+}
\end{pmatrix}
\, ,
\label{matpotential}
\end{eqnarray}
in the basis given in Table~\ref{table_qnumbers} in the same ordering.
The $\pi$ exchange potential between a heavy meson and 
a nucleon is obtained by
\begin{equation}
 V^{\pi}_{1/2^+} = \frac{g_{\pi} g_{\pi NN}}{\sqrt{2}m_N f_{\pi}}
\frac{1}{3} 
\begin{pmatrix}
 0 & \sqrt{3} C_{m_{\pi}} & -\sqrt{6}T_{m_{\pi}}  \\
\sqrt{3}C_{m_{\pi}} & -2C_{m_{\pi}} & -\sqrt{2} T_{m_{\pi}} \\
-\sqrt{6}T_{m_{\pi}} & -\sqrt{2}T_{m_{\pi}} & C_{m_{\pi}} - 2T_{m_{\pi}}
\end{pmatrix}
\vec{\tau}_{P} \cdot \vec{\tau}_N ,
\label{matpi}
\end{equation}

\begin{equation}
 V^{\pi}_{3/2^+} = \frac{g_{\pi}g_{\pi NN}}{\sqrt{2} m_N f_{\pi}}
\frac{1}{3} 
\begin{pmatrix}
 0 &\sqrt{3}C_{m_\pi} &\displaystyle \sqrt{\frac{3}{5}}T_{m_\pi}
 &\displaystyle -3\sqrt{\frac{3}{5}}T_{m_\pi} \\
 \sqrt{3}C_{m_\pi} &-2C_{m_\pi} &\displaystyle \frac{1}{\sqrt{5}}T_{m_\pi}
 &\displaystyle -\frac{3}{\sqrt{5}}T_{m_\pi} \\
\displaystyle
 \sqrt{\frac{3}{5}}T_{m_\pi}&\displaystyle \frac{1}{\sqrt{5}}T_{m_\pi}&\displaystyle
 C_{m_\pi}+\frac{8}{5}T_{m_\pi}&\displaystyle \frac{6}{5}T_{m_\pi} \\
\displaystyle
 -3\sqrt{\frac{3}{5}}T_{m_\pi}&\displaystyle -\frac{3}{\sqrt{5}}T_{m_\pi}&\displaystyle
 \frac{6}{5}T_{m_\pi}&C_{m_\pi}\displaystyle -\frac{8}{5}T_{m_\pi} \\
\end{pmatrix}
\vec{\tau}_{P} \cdot \vec{\tau}_N \, ,
\end{equation}

\begin{equation}
  V^{\pi}_{5/2^+} = \frac{g_{\pi}g_{\pi NN}}{\sqrt{2} m_N f_{\pi}}
\frac{1}{3} 
\begin{pmatrix}
0 &\displaystyle \frac{3}{5}\sqrt{10}T_{m_\pi} &\sqrt{3}C_{m_\pi} &\displaystyle -2\sqrt{\frac{3}{5}}T_{m_\pi} \\
\displaystyle \frac{3}{5}\sqrt{10}T_{m_\pi} &\displaystyle
 C_{m_\pi}-\frac{2}{5}T_{m_\pi}&\displaystyle \sqrt{\frac{6}{5}}T_{m_\pi}&\displaystyle
 \frac{4}{5}\sqrt{6}T_{m_\pi} \\
\sqrt{3}C_{m_\pi}&\displaystyle \sqrt{\frac{6}{5}}T_{m_\pi}&-2C_{m_\pi}&\displaystyle
 -\frac{2}{\sqrt{5}}T_{m_\pi} \\
\displaystyle -2\sqrt{\frac{3}{5}}T_{m_\pi} &\displaystyle \frac{4}{5}\sqrt{6}T_{m_\pi} &\displaystyle
 -\frac{2}{\sqrt{5}}T_{m_\pi} &\displaystyle C_{m_\pi}+\frac{2}{5}T_{m_\pi}
\end{pmatrix}
\vec{\tau}_{P} \cdot \vec{\tau}_N \, ,
\end{equation}
where $C_m=C(r;m)$,  $T_m=T(r;m)$, and $\vec{\tau_{P}}$ 
and $\vec{\tau}_{N}$ are the isospin matrices for $P(P^{\ast})$ and $N$.
The functions $C(r;m)$ and $T(r;m)$ are given by
\begin{align}
 C(r;m) & = \int \frac{d^3 q}{(2 \pi)^3} \frac{m^2}{\vec{q }\,^2 + m^2}
 e^{i\vec{q}\cdot \vec{r}} F(\Lambda_P,\vec{q}\,)F(\Lambda_N,\vec{q}\,) ,    \label{C} \\ 
T(r;m) S_{12}(\hat{r}) & = \int  \frac{d^3 q}{(2 \pi)^3} 
\frac{-\vec{q}\,^2}{\vec{q}\,^2 + m^2} S_{12}(\hat{q})
 e^{i\vec{q}\cdot \vec{r}} F(\Lambda_P,\vec{q}\,)F(\Lambda_N,\vec{q}\,) ,
\label{T}
\end{align}
with $S_{12}(\hat{x}) = 3(\vec{\sigma}_1 \cdot \hat{x}) (\vec{\sigma}_2
\cdot \hat{x}) -\vec{\sigma}_1 \cdot \vec{\sigma}_2$, and 
$F(\Lambda,\vec{q}\,)$ denotes the form factor given by
\begin{align}
 F_\alpha(\Lambda,\vec{q})&=\frac{\Lambda^2-m^2_\alpha}{\Lambda^2+\left|\vec{q}\,\right|^2}
\end{align}
where $m_\alpha$ and $\vec{q}$ are the mass and three-momentum of the
 incoming meson $\alpha\,(=\pi,\rho,\omega)$.
 The corresponding potentials of the $\rho$ meson exchange 
are given by 
\begin{align}
 V^{\rho}_{1/2^+} = & \frac{g_V g_{\rho NN} \beta }{\sqrt{2}m_{\rho}^2}
  \begin{pmatrix}
   C_{m_{\rho}}& 0 & 0 \\
   0 & C_{m_{\rho}} & 0 \\
   0 & 0 & C_{m_{\rho}}
  \end{pmatrix} 
  \vec{\tau}_P \cdot \vec{\tau}_N   \notag \\
 & +\frac{g_V g_{\rho NN} \lambda (1 + \kappa)}{\sqrt{2} m_N}\frac{1}{3}
 \begin{pmatrix}
  0 & 2\sqrt{3} C_{m_{\rho}} & \sqrt{6}T_{m_{\rho}}  \\
  2\sqrt{3}C_{m_{\rho}} & -4C_{m_{\rho}} & \sqrt{2} T_{m_{\rho}} \\
  \sqrt{6}T_{m_{\rho}} & \sqrt{2}T_{m_{\rho}} & 2C_{m_{\rho}} + 2T_{m_{\rho}}
 \end{pmatrix}
\vec{\tau}_{P} \cdot \vec{\tau}_N ,
\label{matrho}
\end{align}
\begin{align}
 V^{\rho}_{3/2^+} = & \frac{g_V g_{\rho NN} \beta }{\sqrt{2}m_{\rho}^2}
  \begin{pmatrix}
   C_{m_{\rho}}& 0 & 0 &0 \\
   0 & C_{m_{\rho}} & 0 &0\\
   0 & 0 & C_{m_{\rho}} & 0 \\
   0 & 0 & 0 & C_{m_{\rho}} 
  \end{pmatrix}
\vec{\tau}_{P} \cdot \vec{\tau}_N \notag \\
& + \frac{g_V g_{\rho NN} \lambda (1 + \kappa)}{ \sqrt{2}m_N}\frac{1}{3}
\begin{pmatrix}
 0 &  2\sqrt{3}C_{m_{\rho}} &\displaystyle -\sqrt{\frac{3}{5}}T_{m_{\rho}} &\displaystyle 3\sqrt{\frac{3}{5}}T_{m_{\rho}} \\
 2\sqrt{3}C_{m_{\rho}} & -4C_{m_{\rho}} &\displaystyle -\frac{1}{\sqrt{5}}T_{m_{\rho}} &\displaystyle \frac{3}{\sqrt{5}}T_{m_{\rho}} \\
\displaystyle -\sqrt{\frac{3}{5}}T_{m_{\rho}} &\displaystyle
 -\frac{1}{\sqrt{5}}T_{m_{\rho}} &\displaystyle  2C_{m_{\rho}}-\frac{8}{5}T_{m_{\rho}} &\displaystyle -\frac{6}{5}T_{m_{\rho}} \\
\displaystyle 3\sqrt{\frac{3}{5}}T_{m_{\rho}} &\displaystyle \frac{3}{\sqrt{5}}T_{m_{\rho}} &\displaystyle -\frac{6}{5}T_{m_{\rho}} &\displaystyle 2C_{m_{\rho}}+\frac{8}{5}T_{m_{\rho}}
\end{pmatrix}
\vec{\tau}_{P} \cdot \vec{\tau}_N , \notag \\
\label{matrho3/2}
\end{align}
\begin{align}
  V^{\rho}_{5/2^+} = & \frac{g_V g_{\rho NN} \beta }{\sqrt{2}m_{\rho}^2}
  \begin{pmatrix}
   C_{m_{\rho}}& 0 & 0 &0 \\
   0 & C_{m_{\rho}} & 0 &0\\
   0 & 0 & C_{m_{\rho}} & 0 \\
   0 & 0 & 0 & C_{m_{\rho}} 
  \end{pmatrix}
\vec{\tau}_{P} \cdot \vec{\tau}_N \notag \\
& + \frac{g_V g_{\rho NN} \lambda (1 + \kappa)}{ \sqrt{2}m_N}\frac{1}{3}
\begin{pmatrix}
0 &\displaystyle -\frac{3}{5}\sqrt{10}T_{m_\rho} &2\sqrt{3}C_{m_\rho}
 &\displaystyle 2\sqrt{\frac{3}{5}}T_{m_\rho} \\
\displaystyle
 -\frac{3}{5}\sqrt{10}T_{m_\rho}&\displaystyle 2C_{m_\rho}+\frac{2}{5}T_{m_\rho}
 &\displaystyle -\sqrt{\frac{6}{5}}T_{m_\rho} &\displaystyle
 -\frac{4}{5}\sqrt{6}T_{m_\rho} \\
2\sqrt{3}C_{m_\rho} &\displaystyle
 -\sqrt{\frac{6}{5}}T_{m_\rho}&-4C_{m_\rho} &\displaystyle
 \frac{2}{\sqrt{5}}T_{m_\rho} \\
\displaystyle 2\sqrt{\frac{3}{5}}T_{m_\rho} &\displaystyle
 -\frac{4}{5}\sqrt{6}T_{m_\rho} &\displaystyle
 \frac{2}{\sqrt{5}}T_{m_\rho}&\displaystyle 2C_{m_\rho}-\frac{2}{5}T_{m_\rho} \\
\end{pmatrix}
\vec{\tau}_{P} \cdot \vec{\tau}_N . \notag \\
\label{matrho5/2}
\end{align}
The $\omega$ meson exchange potential can be obtained by 
replacing the relevant coupling constants and the mass of 
the exchanged meson, and by removing the isospin factor 
$\vec{\tau}_{P} \cdot \vec{\tau}_N$.
The anomalous coupling $\kappa$ for the $\omega$ meson exchange
potential is set as zero in Eqs.~\eqref{matrho}-\eqref{matrho5/2}.


The kinetic terms are given by
\begin{align}
 K_{1/2^+} &= \mbox{diag} \left( -\frac{1}{2 \tilde{m}_P}\bigtriangleup_1 ,
- \frac{1}{2\tilde{m}_{P^*}} \bigtriangleup_1 +\Delta m_{PP^*},
-\frac{1}{2\tilde{m}_{P^*}} \bigtriangleup_1 +\Delta m_{PP^*} \right)\,
 , \\
K_{3/2^+} &= \mbox{diag} \left( -\frac{1}{2 \tilde{m}_P}\bigtriangleup_1 ,
- \frac{1}{2\tilde{m}_{P^*}} \bigtriangleup_1 +\Delta m_{PP^*},
-\frac{1}{2\tilde{m}_{P^*}} \bigtriangleup_1 +\Delta m_{PP^*}
\right. , \notag \\
& \quad \left.-\frac{1}{2\tilde{m}_{P^*}} \bigtriangleup_3 +\Delta
 m_{PP^*} \right) \, , \\
K_{5/2^+} &= \mbox{diag} \left( -\frac{1}{2 \tilde{m}_P}\bigtriangleup_3 ,
- \frac{1}{2\tilde{m}_{P^*}} \bigtriangleup_1 +\Delta m_{PP^*},
-\frac{1}{2\tilde{m}_{P^*}} \bigtriangleup_3 +\Delta m_{PP^*}
\right. , \notag \\
& \quad \left.-\frac{1}{2\tilde{m}_{P^*}} \bigtriangleup_3 +\Delta
 m_{PP^*} \right) \, ,
\end{align}
for $J^P = 1/2^+$, $3/2^+$ and $5/2^+$, respectively. Here, we define
$\bigtriangleup_l = \partial^2 / \partial r^2 + (2/r)\partial / \partial
r  - l(l+1)/r^2$
and $
\tilde{m}_{P{^{(\ast)}}}
= m_N m_{P{^{(\ast)}}}/(m_N +m_{P{^{(\ast)}}}),$ with $\Delta m_{PP^*} = m_{P^*} -m_P$.
The total Hamiltonian is then given by $H_{J^P} = K_{J^P} + V_{J^P}$.

\clearpage



\begin{thebibliography}{9}
\bibitem{comment}
In the present study, we do not discuss $DN$ and $\bar{B}N$. Because there are many
open thresholds and resonances below $DN$ and $\bar{B}N$ thresholds, we
will need to consider more complicated coupled channel dynamics for $DN$
and $\bar{B}N$.
\end{thebibliography}
\end{document}